\documentclass[twocolumn,epsbox]{jpsj2}

\def\runtitle{
Superconductivity in the CuO double chain of Pr$_{2}$Ba$_{4}$Cu$_{7}$O$_{15-\delta}$ on the basis of Tomonaga-Luttinger liquid theory
}
\def\runauthor{Kazuhiro {\sc Sano},  Yoshiaki {\sc \=Ono} and Yuh {\sc Yamada}}

\newcommand{\simj}{\stackrel{>}{_\sim}}
\newcommand{\simk}{\stackrel{<}{_\sim}}

\title{%
Superconductivity in the CuO double chain of Pr$_{2}$Ba$_{4}$Cu$_{7}$O$_{15-\delta}$ \\ on the basis of Tomonaga-Luttinger liquid theory
}

\author{%
Kazuhiro {\sc Sano}\thanks{E-mail address: sano@phen.mie-u.ac.jp}, Yoshiaki {\sc \=Ono}\raisebox{0.5ex}{1,2}  and  Yuh {\sc Yamada}\raisebox{0.5ex}{1,2}
}

\inst{%
Department of Physics Engineering,  Mie University, Tsu, Mie 514-8507       \\ 
\raisebox{0.5ex}{1}Department of Physics, Niigata University, Ikarashi, Niigata 950-2181    \\
\raisebox{0.5ex}{2}Center for Transdisciplinary Research, Niigata University, Ikarashi, Niigata 950-2181
}

\recdate{\today}

\abst{%
Recently, Matsukawa {\it et al.}  have  discovered a new superconductor Pr$_{2}$Ba$_{4}$Cu$_{7}$O$_{15-\delta}$ in which metallic CuO double chains are responsible for the superconductivity. To investigate the superconductivity, we employ the $d$-$p$ double chain model where the tight-binding parameters are determined so as to fit the LDA band structure. On the basis of the Tomonaga-Luttinger liquid theory, we obtain the phase diagram including the superconducting phase in the weak coupling limit. We also calculate the Luttinger liquid parameter $K_{\rho}$ as a function of the electron density $n$ by using the Hartree-Fock approximation. With increasing $n$ from quarter filling, $K_{\rho}$ increases, and then exceeds 1/2 when the superconducting  correlation becomes most dominant.  $K_{\rho}$ has a maximum at an optimal density between quarter- and half-filling. These results are consistent with the experimental observation.
}

\kword{%
Pr$_{2}$Ba$_{4}$Cu$_{7}$O$_{15-\delta}$, CuO double chain, Superconductivity, Tomonaga-Luttinger liquid, $d$-$p$ model
}

\begin{document}
\sloppy
\maketitle


Low-dimensional  strongly correlated electron systems  have  attracted much interest due to the possible relevance to the high-$T_c$ superconductivity.
Although the two-dimensional CuO$_2$ planes play essential roles for the superconductivity, one-dimensional (1D) CuO chains included in some families of high-$T_c$  cuprates have also provided an interesting testing ground for anomalous metallic states due to effects of the electron-correlation \cite{MTakano,Ishida,Barnes}. 

PrBa$_2$Cu$_4$O$_8$ (Pr124) is an excellent system to study the properties of 1D CuO chains, because the electronic conduction in CuO$_2$ plane is suppressed due to the so-called Fehrenbacher-Rice state formed by the strong hybridization between Pr $4f$ and O $2p$ orbitals\cite{Rice}. By studying the anisotropy in the resistivity of a single crystal, it was clarified that the metallic conductivity is caused by conduction in CuO double chains \cite{Horii}. Some researchers suggested that the CuO double chain might be in the Tomonaga-Luttinger liquid state \cite{Mizokawa}.

Pr$_2$Ba$_4$Cu$_7$O$_{15-\delta}$ (Pr247) has both CuO double chains and CuO single chains and shows metallic conductivity at low temperatures owing to the metallic conduction in the CuO double chains. Furthermore, the carrier density can be varied by controlling the amount of oxygen deficiency in the CuO single chains. In the recent reports, the resistivity in Pr247 was investigated by changing the oxygen content, and superconductivity with $T_c \sim 15$K was found after a reduction treatment \cite{Matsukawa, Yamada}. The NQR experiment revealed that the superconductivity is realized at the CuO double chains \cite{Sasaki}. These experiments suggest that the material seems to show the possibility of novel 1D superconductivity.

Many theoretical works have been performed on 1D strongly correlated electron systems confined to a double chain such as the two-chain model and the Ladder model \cite{Dagotto1,Troyer1,Sanotjlad,Sanohublad1,Sanohublad2,Kuroki,Daul,Yamaji,Koike1,Koike2,Seo,Nishimoto}. At half-filling, the system can be described by a Heisenberg model whose ground state is a spin liquid insulator with a finite gap in spin excitation \cite{Dagotto2}. Away from  half-filling, the system becomes a metallic state which maintains a spin gap \cite{Dagotto1,Troyer1,Sanotjlad}. This behavior is explained by the existence of electron pairs  produced  by the dominant fluctuations of the $4k_F$ charge density wave or the interchain-paring fluctuations. The paring fluctuation is expected to dominate over the other fluctuations in the weak coupling regime of the Hubbard model as well as in the strong coupling regime of the $t-J$ model.

\begin{figure}[b]
  \begin{center}
\includegraphics[width=6.3cm]{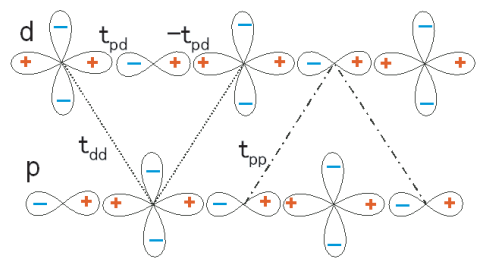}
\end{center}

  \caption[]{
  Schematic diagram of  Cu($3d_{x^2-y^2}$) orbitals and O($2p_{\sigma}$) orbitals in a CuO double chain. 
}
  \label{modelfig}
\end{figure}

Most of the theoretical works on the double chain systems have been done by using simplified single-band models such as the Hubbard model or the  $t-J$ model. As for the single chain systems, two-band models such as the $d$-$p$ model, simulating a CuO linear chain, have also been extensively investigated \cite{Sanodp1,Sudbo,Sanodp2,Sanodp3,Sanodp4,Sanodp5}. They show that the charge fluctuation between $d$ and $p$ orbitals, which is not taken into account in the single-band models, has crucial effect on the superconductivity. In the case of CuO double chain (see Fig. \ref{model}), in addition to such effect, the number of the Fermi points and the values of the corresponding Fermi velocity play crucial roles for the superconductivity.
They directly depend on parameters included in the $d$-$p$ double chain model. Therefore, we need theoretical studies on the basis of a realistic $d$-$p$ double chain model. 

In the present paper, we investigate the $d$-$p$ double chain model, simulating a CuO double chain composed of Cu$(3d_{x^2-y^2})$ orbitals and O$(2p_{\sigma})$ orbitals as shown in Fig. \ref{modelfig}. The tight-binding parameters, {\it i.e.} the hopping integrals and the charge transfer energy are estimated to fit the LDA band structure. By using the weak coupling theory \cite{Balentz,Fabrizio,Emery}, we obtain the phase diagram including the superconducting phase. 

We also calculate the Luttinger liquid parameter $K_{\rho}$ \cite{Haldane,Voit} as a function of the electron density for a finite Coulomb interaction between $d$ electrons $U_d$ by within the Hartree-Fock (HF) approximation.

The Hamiltonian of the $d$-$p$ double chain model is given by 
\begin{eqnarray} 
  H&=& t_{pd}\sum_{i,\sigma} (p_{i\sigma}^{\dagger} d_{i+1\sigma}+h.c.)
    + U_d\sum_{i}\hat{n}_{di\uparrow}\hat{n}_{di\downarrow}
 \nonumber \\
  &+& t_{pp}\sum_{i,\sigma} (p_{i\sigma}^{\dagger} p_{i+1\sigma}+h.c.) 
  + \epsilon_{p}\sum_{i,\sigma} p_{i\sigma}^{\dagger}p_{i\sigma}
 \nonumber \\
   &+& t_{dd}\sum_{i,\sigma} (d_{i\sigma}^{\dagger} d_{i+1\sigma}+h.c.) 
  + \epsilon_{d}\sum_{i,\sigma} d_{i\sigma}^{\dagger}d_{i\sigma},    
  \label{model}
\end{eqnarray} 
where $d^{\dagger}_{i\sigma}$ and $p^{\dagger}_{i\sigma}$ stand for creation operators for a electron with spin $\sigma$ in  a Cu$(3d_{x^2-y^2})$ orbital  at site $i$ and for a hole with spin $\sigma$ in a O$(2p_{\sigma})$ orbital at site $i$, respectively, and   $\hat{n}_{di\sigma}=d_{i\sigma}^{\dagger}d_{i\sigma}$.
 Here,  $t_{pd}$ is the hopping energy  between the nearest-neighbor $d$  and $p$ sites and $t_{pp}$ ($t_{dd}$) is the hopping  energy  between the nearest-neighbor $p$ ($d$) sites. 
 The atomic energy  levels for $p$ and $d$ orbitals are given by $\epsilon_{p}$ and $\epsilon_{d}$, respectively. The charge-transfer energy $\Delta$ is defined as $\Delta=\epsilon_{d}-\epsilon_{p}$. $U_d$ is the on-site Coulomb interaction between $d$ electrons.

\begin{figure}[t]
  \begin{center}
\includegraphics[width=6.8cm]{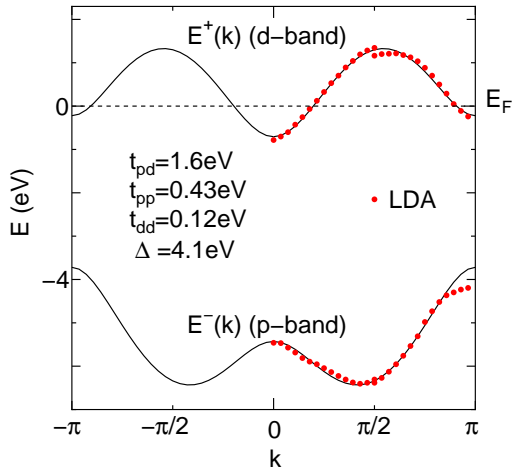}
\end{center}

  \caption[]{
  A typical energy dispersion relation for the CuO double chain.  Solid lines are the tight-binding result $E^{\pm}(k)$  with $t_{pd}=1.6 {\rm eV}$,  $t_{pp}= 0.43 {\rm eV}$,  $t_{dd}=  0.12 {\rm eV}$ and $\Delta = 4.1 {\rm eV}$. Closed circles are the LDA result for Y124 \cite{Draxl}. Dotted line indicates the Fermi level  with $n=0.57$.
}
  \label{bandfig}
\end{figure}

In the noninteracting case with $U_d=0$, the Hamiltonian  eq. (\ref{model}) is easily diagonalized. It yields a dispersion relation:  
$
E^{\pm}(k)=\frac{1}{2} \{ \epsilon_{d}+\epsilon_{p}+2(t_{dd}+t_{pp})\cos{k} \pm \sqrt{(\Delta+2(t_{dd}-t_{pp})\cos{k})^2+16(t_{pd}\cos{(k/2)})^2} \}, 
$
where $k$ is a wave vector and $ E^+(k)$ $( E^-(k) )$ represents the upper (lower) band energy.  
A typical energy dispersion relation $ E^{\pm}(k)$ is depicted in Fig. \ref{bandfig}. As the charge transfer energy is positive, $\Delta=\epsilon_d-\epsilon_p>0$, the upper band $E^{+}(k)$ corresponds to the $d$ band, while the lower band $E^{-}(k)$ corresponds to the $p$ band. 

We estimate the tight-binding parameters in the $d$-$p$ model eq. (\ref{model}) so as to fit the energy dispersion $ E^{\pm}(k)$ to the LDA band structure. 
Since the LDA results for Pr247 have not been available so far, we employ the LDA result for YaBa$_{2}$Cu$_{4}$O$_{8}$ (Y124) \cite{Draxl} in which CuO double chains are included with the same lattice structure as those in Pr247 \cite{Matsukawa}. 
Therefore, the energy dispersions corresponding to the CuO double chain for the both compounds are expected to be almost equivalent. 
Using the least squares method, we obtain the parameters as  $t_{pd}\simeq 1.6 {\rm eV}$,  $t_{pp}\simeq  0.43 {\rm eV}$,  $t_{dd}\simeq  0.12 {\rm eV}$ and $\Delta \simeq 4.1 {\rm eV}$, respectively.
Here, the fitting weight of the $d$ band is taken to be one hundred times as large as that of the $p$ band. This choice gives  better fitting near $E_{\rm F}$ than the fitting with equal weight  for both bands.

These estimated values are  of the same orders of the well known values for the corresponding energies in the CuO$_2$ plane \cite{Hybertsen}. 
The  hopping energies satisfy the relations: $t_{pd}>t_{pp}>t_{dd}>0$, which are consistent with the values of the overlap integrals between atomic orbitals (see Fig. \ref{modelfig}). 
Then, we expect that the obtained tight-binding parameters are realistic to describe the electronic state of the CuO double chain in Pr247.

Since the $d$ band dispersion relation $E^+(k)$ has a double minimum (maximum) structure as shown in Fig. \ref{bandfig}, the number of the Fermi points depends on the  electron density $n$ in the $d$ band of the CuO double chain. There are two Fermi points for small $n$, while, four  Fermi points for large $n$. 
This difference plays crucial role for the superconductivity as will be discussed later. 

The LDA calculation for Y124 \cite{Draxl} has predicted that there are four Fermi points in the $d$ band of the CuO double chain with the electron density $n \simeq 0.57$. As for the case with Pr247, it is considered to be nearly quarter filling $n \approx 0.5$ due to  the Fehrenbacher-Rice effect  mentioned before \cite{Rice}. Then, the CuO double chain of Pr247 is in the boundary between the two Fermi point system and the marginal four Fermi point system as seen in Fig. \ref{bandfig} where the Fermi level for $n=0.57$ is plotted.

On the basis of the Tomonaga-Luttinger liquid theory \cite{Balentz,Fabrizio,Emery,Haldane,Voit}, various types of correlation functions show power-low dependence with critical exponents. These exponents are determined by a single parameter $K_\rho$ in the model which is isotropic in spin space. 
For the single-band model with two Fermi points, $\pm k_{F}$, the SC correlation function decays as $\sim r^{-(1+\frac{1}{K_{\rho}})}$, while the CDW and SDW correlation functions decay as $\sim \cos({2k_F r}) r^{-(1+K_{\rho})}$.
When the system is in the Tomonaga-Luttinger regime,  both of the charge and spin excitations are gapless (we label this regime as $c1s1$). 
In this case, the SC correlation  is dominant for $K_{\rho}>1$, while the CDW or SDW correlation is dominant for $K_{\rho}<1$. 

In contrast to the single-band model, the situation is rather complicated in the two-band model with four Fermi points, $\pm k_{F_1}$ and $\pm k_{F_2}$. 
When the ratio of the two Fermi velocities $|v_{F_1}/v_{F_2}|$ is smaller than a critical value $\sim 8.6$, the low-energy excitations are given by a single gapless charge mode with a gapped spin mode (labeled as $c1s0$) \cite{Balentz,Fabrizio,Emery}. 
In this case, the SC  and the CDW correlations decay as $\sim r^{-\frac{1}{2K_{\rho}}}$ and $\sim \cos[{2(k_{F_2}-k_{F_1}) r}] r^{-2K_{\rho}}$, respectively, while the SDW correlation decays exponentially. 
Hence, the SC correlation is dominant for  $K_\rho >0.5$, while, the CDW correlation is dominant for $K_\rho <0.5$. 
When  the ratio  $|v_{F_1}/v_{F_2}|$  is larger  than the critical value $\sim 8.6$, the low energy excitations are given by two gapless charge modes and two gapless spin modes (labeled as $c2s2$). 

\begin{figure}[t]
  \begin{center}
\includegraphics[width=7.2cm]{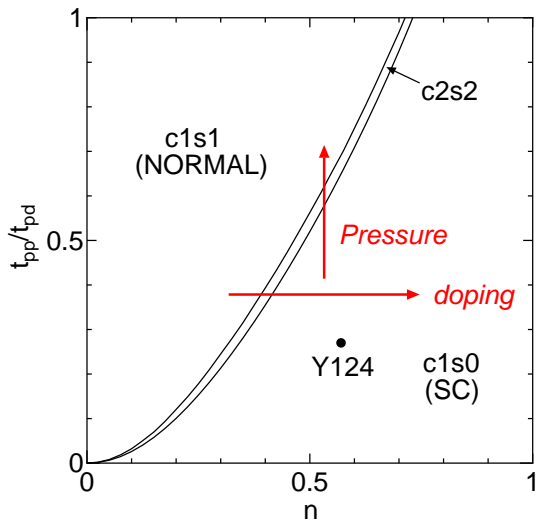}
\end{center}

  \caption[]{
Phase diagram  on the $n$-$t_{pp}$ plane  in the weak coupling limit for $t_{pd}=1.6 {\rm eV}$, $\Delta=4.1 {\rm eV}$ and $t_{dd}/t_{pp}=0.28$. 
Closed circle stands for the parameter point corresponding to the LDA calculation for Y124.
}
  \label{weak1}
\end{figure}

In the non-interacting case with $U_d=0$, the Luttinger liquid parameter $K_{\rho}$ is always unity. In the weak coupling limit $U_d \to 0$, {\it i.e.}, in the limit $K_{\rho} \to 1$, the SC correlation is most dominant in the $c1s0$ region (SC phase), while, the CDW or SDW correlation is most dominant in the $c1s1$ and $c2s2$ regions (normal phases) \cite{Balentz,Fabrizio,Emery}. In Fig. \ref{weak1}, we plot the phase diagram in the weak coupling limit on the $n$-$t_{pp}$ plane. The values of $t_{pd}=1.6 {\rm eV}$ and $\Delta=4.1 {\rm eV}$ together with the ratio $t_{dd}/t_{pp}=0.28$ are fixed to the corresponding values in Fig. \ref{bandfig}.

As mentioned before, the CuO double chain of Pr247 is expected to be in the boundary region between the $c1s1$ and the $c1s0$ phases. When the system is initially in the $c1s1$ phase, the electron doping effect brings about the phase transition from the $c1s1$ to the $c1s0$ (see Fig.  \ref{weak1}). This is consistent with the experimental observation in Pr247, where the superconductivity is caused by the oxygen reduction corresponding to the electron doping effect in the CuO double chain \cite{Matsukawa,Yamada}. 

When the distance between the two chains of a CuO double chain decreases, the hopping terms $t_{pp}$ and $t_{dd}$ are expected to increase. Therefore, the pressure effect might lead to the phase transition from the $c1s0$ to the $c1s1$ as shown in Fig. \ref{weak1}. This is again consistent with the recent experiment in Pr247 under high pressure \cite{Yamada2}, where the superconductivity is suppressed and finally disappears due to the pressure effect.

Now, we calculate the Luttinger liquid parameter $K_{\rho}$ for finite Coulomb interaction $U_d$. 
Based on the Tomonaga-Luttinger liquid theory, $K_{\rho}$ is related to  the charge susceptibility $\chi_c$ and the Drude weight $D$ by the following equation: \cite{Voit}
\begin{equation}
      K\sb{\rho}=\frac{1}{2}(\pi \chi_c D)^{1/2} 
 \label{Krho}
\end{equation}
with 
\begin{equation}
   \chi_c^{-1}= \frac{\partial^2 E_g}{\partial n^2}, \ \ \ \ 
   D=\pi \frac{\partial^2 E_g}{\partial \phi^2}, 
 \label{Drude}
\end{equation}
where $E_g$ is the ground state energy per unit cell as a function of the electron density $n$ and the magnetic flux $\phi$. 
To calculate $K_{\rho}$ in the $d$-$p$ model eq.(\ref{model}), we estimate $E_g$ within the HF approximation and substitute it into eq.(\ref{Drude}). The obtained value of $K_{\rho}$ is valid up to the first order of $U_d$ and is expected to be a good approximation in the weak and intermediate coupling regimes as shown previous works \cite{Sanohublad1,Sanohublad2,Sanodp4,Sanodp5}.

\begin{figure}[t]
  \begin{center}
\includegraphics[width=7.0cm]{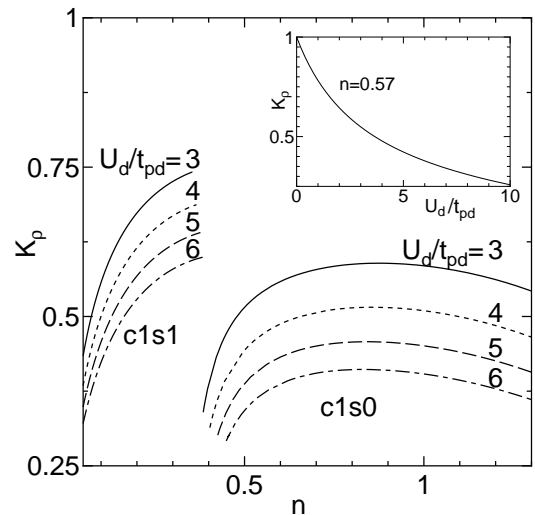}
\end{center}

  \caption[]{
The Luttinger parameter $K_{\rho}$  as a function of the electron density $n$ for several $U_d$ for $t_{pd}=1.6 {\rm eV}$,  $t_{pp}= 0.43 {\rm eV}$,  $t_{dd}=  0.12 {\rm eV}$ and $\tilde{\Delta} = 4.1 {\rm eV}$. Inset shows $K_{\rho}$ as a function of $U_d$ for $n=0.57$. 
}
  \label{Krow}
\end{figure}

Fig. \ref{Krow} shows $K_{\rho}$  as a function of $n$ for several values of $U_d/t_{pd}$ with $t_{pd}=1.6 {\rm eV}$,  $t_{pp}= 0.43 {\rm eV}$,  $t_{dd}=  0.12 {\rm eV}$ and $\tilde{\Delta}|_{n=0.57} = 4.1 {\rm eV}$, where $\tilde{\Delta}$ is the renormalized charge-transfer energy within the HF approximation and is explicitly given by $\tilde{\Delta} = \Delta+\frac{U_d n_d}{2}$. 
As $\tilde{\Delta}$ depends on $n$ and $U_d$, we determine the bare charge transfer energy $\Delta$ so as to fix $\tilde{\Delta}$ to the corresponding value in Fig. \ref{bandfig}: $\tilde{\Delta}=4.1{\rm eV}$ at $n=0.57$ for each $U_d$. 

As shown in Fig. \ref{Krow}, the $c1s1$  phase appears  for $ n\simk 0.4$, while the $c1s0$ phase  appears for $ n\simj 0.4$. In $c1s0$ phase, $K_{\rho}$ increases with increasing $n$ and then have a maximum at an optimal electron density $n \simeq 0.85$. In the $c1s0$ phase with $U_d/t_{pd} \simk 4$, the value of $K_{\rho}$ exceeds $1/2$ for a region of $n$ around $0.85$, where the SC correlation becomes most dominant as compared with the other correlations (SC phase).  On the other hand, in the $c1s1$ phase, $K_{\rho}$ is always smaller than unity and the CDW or SDW correlation is most dominant (normal phase).   
As shown in the inset in Fig. \ref{Krow}, $K_{\rho}$ decreases monotonically with increasing  $U_d$; the similar behavior has been observed in the $U$-dependence of $K_{\rho}$ in the Hubbard ladder model \cite{Sanohublad1,Sanohublad2}. 

Experimentally, the superconductivity of Pr247 is observed only for a range of the oxygen reduction rate $\delta \simj 0.3$ and $T_c$ shows a maximum at an optimal reduction rate $\delta \sim 0.45$ \cite{Yamada}. When we assume the electron density of the CuO double chain in Pr247 with $\delta=0$ to be $n\simeq 0.4$, we can guess $n \simeq 0.4+\delta$ for finite $\delta$. Then, for $U_d/t_{pd}=4$ corresponding to $U_d=6.4$eV, the SC region ($n\simj 0.7$) with $K_{\rho}>1/2$ as well as the optimal electron density ($n\simeq 0.85$) with a maximum in $K_{\rho}$ seems to be consistent with the experimental observation mentioned above.


To summarize, we have investigated the superconductivity in the $d$-$p$ double chain model, simulating a CuO double chain of Pr247, where the tight-binding parameters are determined so as to fit the LDA band structure. 
On the basis of the Tomonaga-Luttinger liquid theory, we have obtained the phase diagram in the weak coupling limit, which provides us a clear understanding of the doping and the pressure dependence of the superconductivity in Pr247 \cite{Yamada,Yamada2}. 

The Luttinger liquid parameter $K_{\rho}$ has also been obtained as a function of the electron density $n$ using the Luttinger-liquid relation combined with the HF approximation. The doping dependence of $K_{\rho}$ is in good agreement with that of $T_c$ in Pr247 \cite{Yamada}, when we assume that $T_c$ is finite for $K_{\rho}>1/2$ in the $c1s0$ phase and is monotonically increasing function of $K_{\rho}$. Although the finite value of $T_c$ is not obtained in purely one-dimensional systems, an approximate value of $T_c$ could be estimated as a function of $K_{\rho}$ by taking into account of a three-dimensionality due to the effect of couplings between the double chains. We will report it in a subsequent paper. 
Even in the normal state above $T_c$, a kind of transition between the $c1s1$ phase and the $c1s0$ phase is expected. In fact, such a transition has been observed in the doping and/or the pressure dependence of transport properties such as the resistivity and the Hall coefficient \cite{Yamada,Yamada2}. In addition to the transport properties, a spin gap is expected to exist in the $c1s0$ phase. A spin gap like behavior has been observed in the recent NQR experiment \cite{Sasaki,Sasaki2}, where $(T_1 T)^{-1}$ is suppressed in the superconducting sample as compared with the non-superconducting sample even above $T_c$. 

Taking account of the LDA calculation and our result, we can expect that the CuO double chain of Y124 is also in the $c1s0$ phase and shows the superconductivity as well as Pr247.
We note that  a rapid decrease in the resistivity below $15$K has been observed in YBa$_2$(Cu$_{1-x}$Zn$_x$)$_4$O$_8$ with high Zn doping $0.05<x<0.1$ \cite{Miyatake}. In this case, the superconductivity in the CuO$_2$ plane is considered to disappear, but  the CuO double chain might preserve the superconductivity. To be more conclusive, we need further investigation in this regard.

%

The authors thank A. Matsushita and S. Sasaki for many useful discussions. 
This work was partially supported by the Grant-in-Aid for Scientific Research from the Ministry of Education, Culture,  Sports, Science  and Technology.

\end{document}